\documentclass[preprint,12pt    ]{elsarticle}
\usepackage{graphicx}
\usepackage{float}
\usepackage{latexsym,amsbsy,epsfig,fancybox}
\usepackage{amstext}
\usepackage{subfigure}
\usepackage{amsfonts,textgreek}
\usepackage{mathrsfs}
\usepackage{algorithm,algorithmic}
\usepackage{mathtools}
\usepackage{tikz}
\usetikzlibrary{positioning,arrows.meta,shapes.multipart}
\tikzset{
  block/.style={rectangle, draw, thick, rounded corners, minimum width=2cm, minimum height=1cm, align=center},
  smallblock/.style={rectangle, draw, thick, minimum width=1.2cm, minimum height=0.8cm, align=center},
  sum/.style={circle, draw, thick, minimum size=6mm, inner sep=0pt},
  mult/.style={circle, draw, thick, minimum size=6mm, inner sep=0pt},
  >=Stealth
}
\usepackage{pstricks}
\usepackage{array}
\usepackage{amssymb}
\usepackage{multirow}
\usepackage{booktabs}
\usepackage{verbatim}
\usepackage{natbib}
\biboptions{sort&compress}
\usepackage{url}
\usepackage{xr}
\usepackage[margin=3cm]{geometry}
\usepackage{soul}
\usepackage[colorlinks=true,citecolor=blue]{hyperref}

\newcommand{\ac}[1]{\textcolor{red}{add citation}}

\newcommand{\af}[1]{\textcolor{red}{add figure}}
\newcommand{\ar}[1]{\textcolor{red}{add crossref}}
\tikzstyle{nicebox}=[draw=black!100, fill=white!10, rectangle, inner sep=4pt, inner ysep=16pt]
\tikzstyle{niceboxtitle}=[draw=black!100, fill=white, text=black, rectangle]

\usetikzlibrary{arrows,decorations.markings}

\journal{arXiv}
\begin{document}
\begin{frontmatter}

\title{StockBot 2.0: Vanilla LSTMs Outperform Transformer-based Forecasting for Stock Prices}

\author[1]{Shaswat Mohanty\corref{cor1}} 
\ead{shaswatm@stanford.edu}

\cortext[cor1]{Corresponding author}
\address[1]{Department of Mechanical Engineering, Stanford University, CA 94305-4040, USA}
\begin{abstract}
Accurate forecasting of financial markets remains a long-standing challenge due to complex temporal and often latent dependencies, non-linear dynamics, and high volatility. Building on our earlier recurrent neural network framework, we present an enhanced StockBot architecture that systematically evaluates modern attention-based, convolutional, and recurrent time-series forecasting models within a unified experimental setting. While attention-based and transformer-inspired models offer increased modeling flexibility, extensive empirical evaluation reveals that a carefully constructed vanilla LSTM consistently achieves superior predictive accuracy and more stable buy/sell decision-making when trained under a common set of default hyperparameters. These results highlight the robustness and data efficiency of recurrent sequence models for financial time-series forecasting, particularly in the absence of extensive hyperparameter tuning or the availability of sufficient data when discretized to single-day intervals. Additionally, these results underscore the importance of architectural inductive bias in data-limited market prediction tasks.
\end{abstract}

\begin{keyword}
Stock Prediction \sep StockBot \sep Transformer \sep Time Series Forecasting \sep Attention
\end{keyword}

\end{frontmatter}
\section{Introduction}
\label{sec:Intro}

Predicting stock price movements remains a central challenge in quantitative finance due to the inherent non-linearity, noise, and structural volatility of financial time-series data. Traditional statistical and econometric models often rely on linear assumptions and struggle to capture the dynamic dependencies that govern market behavior~\cite{murphy1999technical, tsai2010combining}. Deep learning methods, particularly recurrent neural networks (RNNs) and their variants such as long short-term memory (LSTM) networks, have demonstrated improved performance in modeling temporal dependencies and nonlinear correlations~\cite{di2016artificial, fischer2018deep, selvin2017stock}. \cite{mohanty2022stockbot} has presented the benefits of stacking multiple LSTMs through their ablative studies, demonstrating superior predictive performance of stock prices, especially within the energy sector.  A large corpus of such research exists in the literature and has shown promise in predicting stock price movements, but they fall short in capturing long-term estimations due to hysteresis~\cite{chen2023challenges} as well as high volatility index (VIX) fluctuations in the short-term~\cite{finnegan2024less}.

Subsequent advances in sequence modeling have introduced alternative architectures such as transformers~\cite{vaswani2017attention} and temporal convolutional networks (TCNs)~\cite{lea2017temporal}, which enable more efficient and expressive representations of long-range dependencies. Transformers, through their self-attention mechanism, can flexibly model contextual relevance across time, while TCNs leverage dilated causal convolutions to achieve comparable receptive fields with fewer parameters. These architectures have shown strong generalization across diverse time-series applications, motivating their adaptation for financial forecasting~\cite{liu2019time, zhang2022transformer, wang2022stock}.

In this work, we extend our previous StockBot framework by incorporating a diverse family of temporal forecasting models, including recurrent, transformer-inspired attention-based, convolutional, and hybrid attention--recurrent architectures, within a unified evaluation setting. This design enables a controlled comparison of model expressiveness, generalization behavior, and downstream trading interpretability across multiple market sectors. By evaluating all architectures under identical training protocols and default hyperparameter settings, our analysis emphasizes the role of architectural inductive bias in shaping predictive performance and decision stability, as assessed through a simple buy/sell trading strategy.

The remainder of this paper is organized as follows. Section~\ref{sec:data_methods} describes the dataset construction, normalization procedures, and model architectures. Section~\ref{sec:results} presents comparative results across model families, along with an evaluation of StockBot’s trading performance. Finally, Section~\ref{sec:conclusions} summarizes key findings and outlines directions for future work.

\section{Data and Methods}
\label{sec:data_methods}

\subsection{Data Preparation}

The dataset and preprocessing steps follow the procedure established in our previous work~\cite{mohanty2022stockbot}. Historical stock price data are obtained from Yahoo! Finance~\cite{Yahoofin}, focusing on equities listed on the NYSE and NASDAQ. The adjusted closing price is used as the primary variable for prediction.

A sliding window approach constructs supervised learning sequences. For a given \texttt{past$\_$history} of $k$ days, the model predicts the stock price for the subsequent \texttt{forward$\_$look} days. While most experiments use \texttt{forward$\_$look}$=1$, the framework generalizes to multi-step prediction. The data span from 2010 - 2020 and are split into $80\%$ training and $20\%$ testing subsets, with temporal ordering preserved. All series are normalized using z-score scaling based on the training set statistics to prevent data leakage. This data selection and preprocessing are consistent with the earlier evaluation in \citep{mohanty2022stockbot}.

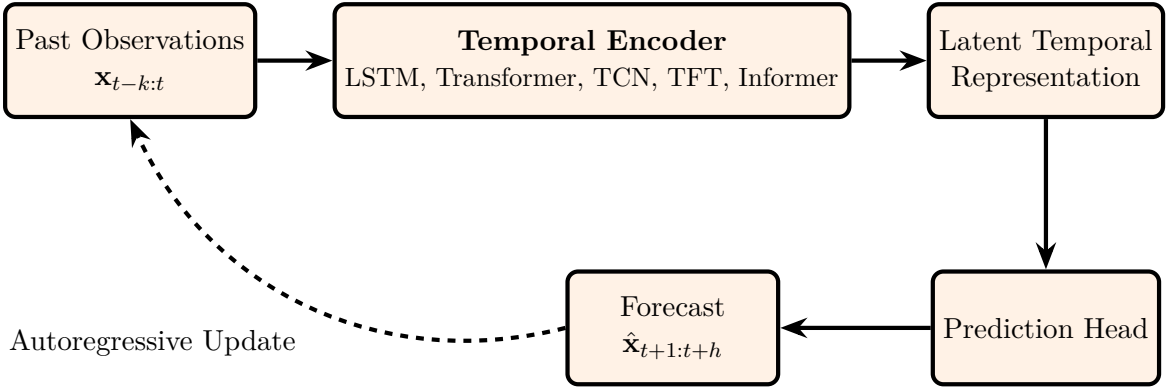
\begin{figure}[!htb]
\centering
\hspace{-0em}
\begin{tikzpicture}[
    node distance=2.0cm,
    every node/.style={font=\small},
    block/.style={
        draw,
        rectangle,
        rounded corners,
        ultra thick,
        fill=orange!10,
        minimum height=1.5cm,
        minimum width=2.8cm,
        align=center
    },
    arrow/.style={->, ultra thick}
]

\node[block] (input) {Past Observations\\
$\mathbf{x}_{t-k:t}$};

\node[block, right=1cm of input] (encoder) {\textbf{Temporal Encoder}\\
{\footnotesize{LSTM, Transformer, TCN, TFT, Informer}}};

\node[block, right=1cm of encoder] (latent) {Latent Temporal\\
Representation};

\node[block, below=of latent] (head) {Prediction Head};

\node[block, left=of head] (output) {Forecast\\
$\hat{\mathbf{x}}_{t+1:t+h}$};

\draw[arrow] (input) -- (encoder);
\draw[arrow] (encoder) -- (latent);
\draw[arrow] (latent) -- (head);
\draw[arrow] (head) -- (output);

\draw[arrow, dashed, bend left=40] 
    (output.west) to node[left, xshift=-0em, yshift=-0.5cm]{Autoregressive Update} (input.south);

\end{tikzpicture}
\caption{
Generic temporal forecasting framework shared by all models considered in this work.
Past observations are encoded into a latent temporal representation, which is mapped to future predictions.
The dashed loop indicates optional autoregressive forecasting for multi-step prediction.
}
\label{fig:generic_temporal_model}
\end{figure}

\subsection{Model Architectures}

We evaluate a family of temporal deep learning models under a unified framework, each implemented as a subclass of a base \texttt{LSTM$\_$Model}. This enables consistent data handling, training, and evaluation across different model types. All models can perform two types of forecasting at the time of evaluation: (i) autoregressive forecasting by recursively feeding the most recent prediction as the next input. This allows evaluation of stability and drift behavior over multiple prediction horizons, and (ii) teacher-forcing forecasting, where ground-truth values are supplied at each step, providing an upper bound on predictive accuracy in the absence of compounding errors.

\subsubsection{LSTM and Attention-Enhanced LSTM}

The baseline model uses a stacked long short-term memory (LSTM) network, which captures temporal dependencies through gated recurrent units. The forward pass of an LSTM layer at time step $t$ is governed by
Eqs.~(\ref{eq:first})--(\ref{eq:last}):
\begin{align}
    f_t &= \sigma_g(W_f x_t + U_f h_{t-1} + b_f), \label{eq:first}\\
    i_t &= \sigma_g(W_i x_t + U_i h_{t-1} + b_i),\\
    o_t &= \sigma_g(W_o x_t + U_o h_{t-1} + b_o),\\
    \tilde{c}_t &= \sigma_c(W_c x_t + U_c h_{t-1} + b_c),\\
    c_t &= f_t \odot c_{t-1} + i_t \odot \tilde{c}_t,\\
    h_t &= o_t \odot \sigma_h(c_t), \label{eq:last}
\end{align}
where $x_t$ is the input vector to the LSTM unit; $f_t$, $i_t$, and $o_t$
denote the forget, input, and output gate activations; $c_t$ and $h_t$
are the cell and hidden states, respectively. The functions
$\sigma_g(\cdot)$ and $\sigma_c(\cdot)=\sigma_h(\cdot)$ correspond to the
sigmoid and hyperbolic tangent activations. The matrices $W$, $U$, and
biases $b$ are learned during training.

An attention-augmented variant introduces a learnable attention layer over LSTM outputs, enabling adaptive weighting of relevant time steps. Given the sequence of hidden states $\{h_1, h_2, \dots, h_T\}$ produced by
the stacked LSTM layers, a Bahdanau-style attention mechanism is applied to compute a context vector. The query vector is defined as the temporal average of the hidden states,
\begin{equation}
    q = \frac{1}{T} \sum_{t=1}^{T} h_t.
\end{equation}
The attention score for each time step is computed as
\begin{equation}
    e_t = v^\top \tanh\!\left(W_1 h_t + W_2 q\right),
\end{equation}
where $W_1$ and $W_2$ are learnable projection matrices and $v$ is a learnable
attention vector. The normalized attention weights are obtained via
\begin{equation}
    \alpha_t = \frac{\exp(e_t)}{\sum_{k=1}^{T} \exp(e_k)}.
\end{equation}
The resulting context vector is a weighted sum of the hidden states,
\begin{equation}
    c = \sum_{t=1}^{T} \alpha_t h_t.
\end{equation}
Finally, the context vector is mapped to the prediction horizon through a
fully connected layer,
\begin{equation}
    \hat{y} = W_o c + b_o.
\end{equation}

\subsubsection{Temporal Convolutional Network (TCN)}

The Temporal Convolutional Network (TCN) models sequential dependencies using causal one-dimensional convolutions, ensuring that predictions at time $t$ depend only on inputs from time steps $\leq t$. Given an input sequence
$\mathbf{X} = \{x_1, x_2, \dots, x_T\}$ with $x_t \in \mathbb{R}^{d_{\text{in}}}$, the first causal convolutional layer computes
\begin{equation}
\mathbf{H}^{(1)}_t
= \phi\!\left(
\sum_{k=0}^{K-1} W^{(1)}_k \, x_{t-k} + b^{(1)}
\right),
\end{equation}
where $K$ is the kernel size, $\phi(\cdot)$ denotes the ReLU activation, and
$W^{(1)}_k$ and $b^{(1)}$ are the learnable convolutional weights and bias.
A second causal convolutional layer further transforms the sequence:
\begin{equation}
\mathbf{H}^{(2)}_t
= \phi\!\left(
\sum_{k=0}^{K-1} W^{(2)}_k \, \mathbf{H}^{(1)}_{t-k} + b^{(2)}
\right).
\end{equation}

\noindent To obtain a fixed-dimensional representation of the temporal sequence, global
average pooling is applied across time:
\begin{equation}
\bar{\mathbf{h}}
= \frac{1}{T} \sum_{t=1}^{T} \mathbf{H}^{(2)}_t.
\end{equation}

Finally, the pooled representation is mapped to the output prediction via a
fully connected layer:
\begin{equation}
\hat{\mathbf{y}} = \bar{\mathbf{h}} W_o + b_o,
\end{equation}
where $\hat{\mathbf{y}} \in \mathbb{R}^{\texttt{forward\_look}}$ represents the
forecast for the next $\texttt{forward\_look}$ time steps.

\subsubsection{Informer Variant}

These transformer extensions reduce computational cost and enhance temporal representation. Informer leverages probSparse attention for long-horizon forecasting;
The Informer-style model operates on an input time series
$\mathbf{X} = \{x_1, x_2, \dots, x_T\}$, where $x_t \in \mathbb{R}^{d_{\text{in}}}$
and $T$ denotes the length of the historical window. The input is first
embedded into a latent space of dimension $d$:
\begin{equation}
\mathbf{H}^{(0)} = \mathbf{X} W_e + b_e,
\end{equation}
where $W_e \in \mathbb{R}^{d_{\text{in}} \times d}$.
The embedded sequence is processed by a stack of $L$ transformer encoder
blocks. At encoder layer $\ell$, multi-head self-attention is computed as
\begin{align}
\mathbf{Q}_h^{(\ell)} &= \mathbf{H}^{(\ell-1)} W_h^Q, \\
\mathbf{K}_h^{(\ell)} &= \mathbf{H}^{(\ell-1)} W_h^K, \\
\mathbf{V}_h^{(\ell)} &= \mathbf{H}^{(\ell-1)} W_h^V,
\end{align}
for attention head $h$. The scaled dot-product attention is given by
\begin{equation}
\text{Attn}_h^{(\ell)} =
\text{softmax}\!\left(
\frac{\mathbf{Q}_h^{(\ell)} \mathbf{K}_h^{(\ell)\top}}{\sqrt{d_k}}
\right)
\mathbf{V}_h^{(\ell)}.
\end{equation}

\noindent The outputs of all heads are concatenated and linearly projected:
\begin{equation}
\text{MHA}^{(\ell)} =
\text{Concat}\big(\text{Attn}_1^{(\ell)}, \dots, \text{Attn}_H^{(\ell)}\big) W^O.
\end{equation}

\noindent A residual connection and layer normalization are applied:
\begin{equation}
\tilde{\mathbf{H}}^{(\ell)} =
\text{LayerNorm}\!\left(
\mathbf{H}^{(\ell-1)} + \text{MHA}^{(\ell)}
\right).
\end{equation}

\noindent The feed-forward sublayer is defined as
\begin{align}
\text{FFN}(\tilde{\mathbf{H}}^{(\ell)}) &=
\phi\!\left(
\tilde{\mathbf{H}}^{(\ell)} W_1 + b_1
\right) W_2 + b_2, \\
\mathbf{H}^{(\ell)} &=
\text{LayerNorm}\!\left(
\tilde{\mathbf{H}}^{(\ell)} + \text{FFN}(\tilde{\mathbf{H}}^{(\ell)})
\right),
\end{align}
where $\phi(\cdot)$ denotes the GELU activation.
After the final encoder layer, temporal information is aggregated using
global average pooling:
\begin{equation}
\bar{\mathbf{h}} = \frac{1}{T} \sum_{t=1}^{T} \mathbf{H}^{(L)}_t.
\end{equation}

\noindent The final prediction for the next $\texttt{forward\_look}$ time steps is
obtained via a linear projection:
\begin{equation}
\hat{\mathbf{y}} = \bar{\mathbf{h}} W_o + b_o.
\end{equation}

Autoformer introduces series decomposition blocks for trend-seasonality modeling; and FEDformer applies frequency-domain attention for improved efficiency on long sequences.

\subsubsection{Temporal Fusion Transformer (TFT)}

The TFT combines recurrent encoders with multi-head attention and gating mechanisms to handle both static covariates and dynamic temporal inputs. It is particularly suited for multivariate forecasting and interpretability of attention over multiple feature groups.
The Temporal Fusion Transformer (TFT) model combines recurrent encoding,
self-attention, and gated residual connections to model both short-term
temporal dynamics and long-range dependencies. Given an input sequence
$\mathbf{X} = \{x_1, x_2, \dots, x_T\}$ with $x_t \in \mathbb{R}^{d_{\text{in}}}$,
the model first applies a stacked LSTM encoder:
\begin{align}
\mathbf{H}^{(1)} &= \text{LSTM}_1(\mathbf{X}), \\
\mathbf{H}^{(2)} &= \text{LSTM}_2(\mathbf{H}^{(1)}),
\end{align}
where $\mathbf{H}^{(2)} \in \mathbb{R}^{T \times d}$ and $d$ denotes the hidden
dimension.

To capture long-range temporal dependencies, multi-head self-attention is
applied to the LSTM outputs. For attention head $h$, the query, key, and value
matrices are given by
\begin{align}
\mathbf{Q}_h &= \mathbf{H}^{(2)} W_h^Q, \\
\mathbf{K}_h &= \mathbf{H}^{(2)} W_h^K, \\
\mathbf{V}_h &= \mathbf{H}^{(2)} W_h^V.
\end{align}
The scaled dot-product attention is computed as
\begin{equation}
\text{Attn}_h =
\text{softmax}\!\left(
\frac{\mathbf{Q}_h \mathbf{K}_h^\top}{\sqrt{d_k}}
\right)\mathbf{V}_h.
\end{equation}

\noindent The outputs of all attention heads are concatenated and normalized:
\begin{equation}
\mathbf{A} =
\text{LayerNorm}\!\left(
\text{Concat}(\text{Attn}_1, \dots, \text{Attn}_H)
\right).
\end{equation}

\noindent A gating mechanism is then applied to control the information flow from the
attention block:
\begin{align}
\mathbf{G} &= \sigma(\mathbf{A} W_g + b_g), \\
\mathbf{Z} &= \mathbf{G} \odot \mathbf{A},
\end{align}
where $\sigma(\cdot)$ denotes the sigmoid activation and $\odot$ represents
element-wise multiplication.
The gated attention output is combined with the LSTM representation via a
residual connection:
\begin{equation}
\mathbf{H} = \mathbf{H}^{(2)} + \mathbf{Z}.
\end{equation}

\noindent Temporal information is aggregated using global average pooling:
\begin{equation}
\bar{\mathbf{h}} = \frac{1}{T} \sum_{t=1}^{T} \mathbf{H}_t.
\end{equation}

\noindent Finally, the prediction for the next $\texttt{forward\_look}$ time steps is
obtained via a linear output layer:
\begin{equation}
\hat{\mathbf{y}} = \bar{\mathbf{h}} W_o + b_o.
\end{equation}

\subsection{Training and Hyperparameters}

Each model is trained using the Adam optimizer with an initial learning rate of $10^{-3}$. Early stopping based on validation loss is applied to prevent overfitting.

\begin{table}[H]
\centering
\begin{tabular}{|l|c||l|c|}
\hline
Hyperparameter & Value & Hyperparameter & Value \\
\hline
\texttt{past$\_$history} & 60 & Batch size & 64 \\
Epochs & 500 & Dropout & 0.1 \\
Hidden dimension & 64 & Feed-forward dimension & 128 \\
Transformer heads & 4 & Train-test split & $80\%/20\%$ \\
\hline
\end{tabular}
\caption{Default hyperparameters used across all models.}
\label{tab:hyperparams_all}
\end{table}

\subsection{Stock predicting bot}\label{sec:buy_sell_bot}
In this paper, we will continue to assess the efficacy of the forecasting models by evaluating the performance of a bot that executes buy/sell operations at the time of closing every day, thereby maximizing gains, as described earlier in \cite{mohanty2022stockbot}. The bot makes decisions based on our predictions of the stock prices. These decisions are analytically enforced and are not learned separately. The bot decision is made by noting a two-step process: We first document the $\delta_i$ changes which are defined by $\delta_i = \text{sign}(c_{i+1} - c_i)$, where $c_i$ is the stock price on the $i^{\rm th}$ day. We then look at the changes in the $\delta_i$, by tracking $\Delta_i = \delta_{i+1}-\delta_i$. $\Delta = -2$ corresponds to the end of a dip, where we buy whereas $\Delta_i=2$ corresponds to the beginning of a dip where we sell. The algorithm is summarized below:
\begin{algorithm}[H]
Obtain the predicted trajectory, $\mathbf{c}$ (stock forecast for \texttt{forward\_look} days;

Compute the nature of change: $\delta_i = \text{sign}(c_{i+1} - c_i)$ ; 

Compute the curvature of the forecast: $\Delta_i = \delta_{i+1}-\delta_i$;

Make the decision: $\text{Decision} = \begin{cases} \Delta_i = 2 \rightarrow \text{ sell (Indicates local maxima)} \\  \Delta_i = 2 \rightarrow \text{ sell (Indicates no change)} \\ \Delta_i = -2 \rightarrow \text{ sell (Indicates local minima)}. \end{cases}$

 \caption{: StockBot decision making algorithm}
\end{algorithm}

\section{Results}
\label{sec:results}

We evaluate all forecasting architectures within a unified experimental framework, comparing both predictive accuracy and downstream trading performance as evaluated by the StockBot. All models are evaluated under identical default hyperparameters without architecture-specific tuning to ascertain the models that can perform reasonably well out of the box in the absence of resources for extensive hyperparameter optimization.

\subsection{One-Day-Ahead Forecasting}

We first consider the task of predicting one trading day ahead, where the model is recursively applied to generate longer trajectories. Two evaluation modes are examined: autoregressive forecasting, where the model’s own predictions are fed back as inputs, and teacher forcing, where ground-truth prices are provided at each step during inference.

\begin{figure}[H]
\centering
\includegraphics[width=0.95\textwidth]{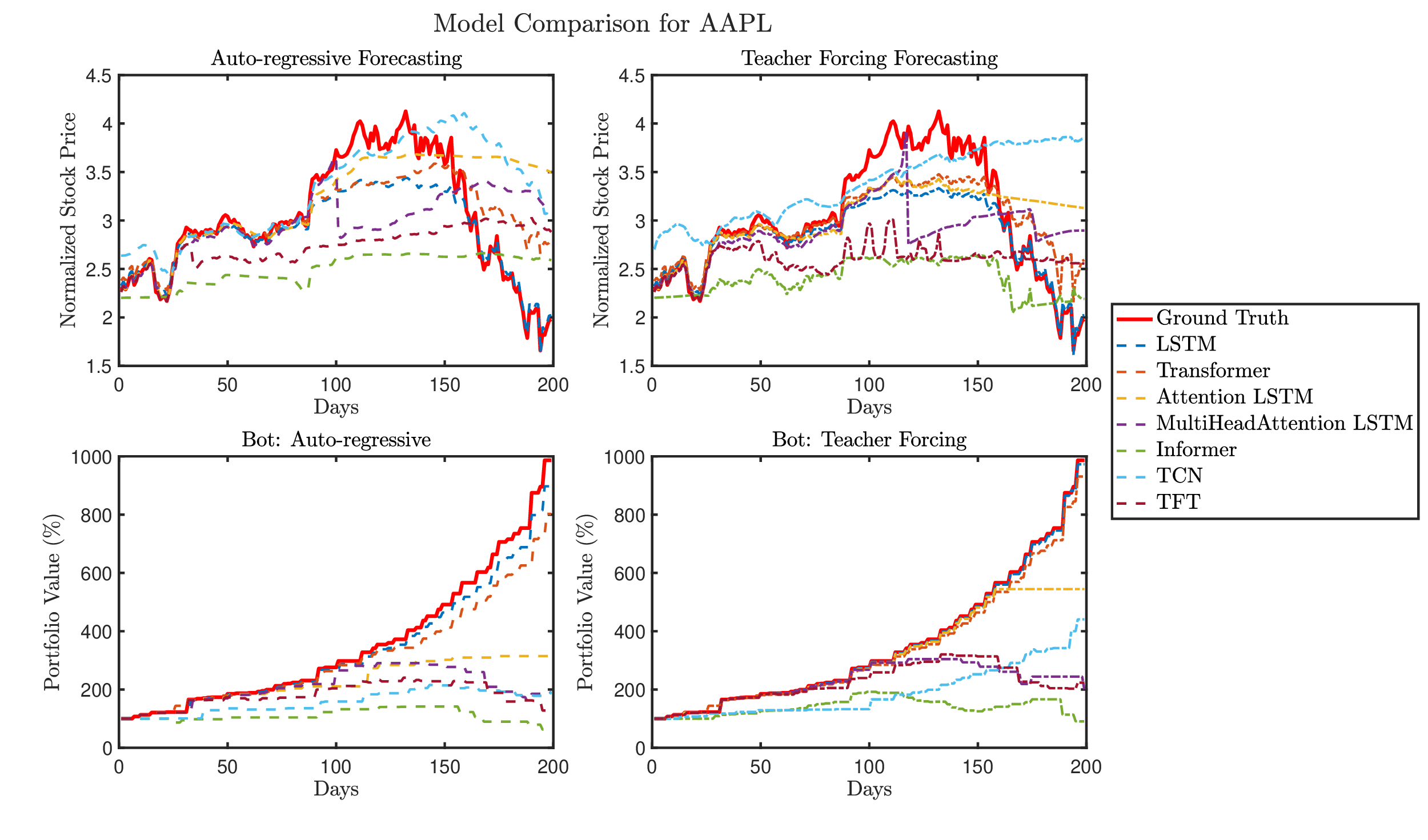}
\caption{One-day-ahead forecasting results for AAPL. Top-left: autoregressive price prediction. Top-right: teacher-forced price prediction. Bottom-left: portfolio value driven by autoregressive forecasts. Bottom-right: portfolio value driven by teacher-forced forecasts.}
\label{fig:aapl_oneday}
\end{figure}

\begin{figure}[H]
\centering
\includegraphics[width=0.95\textwidth]{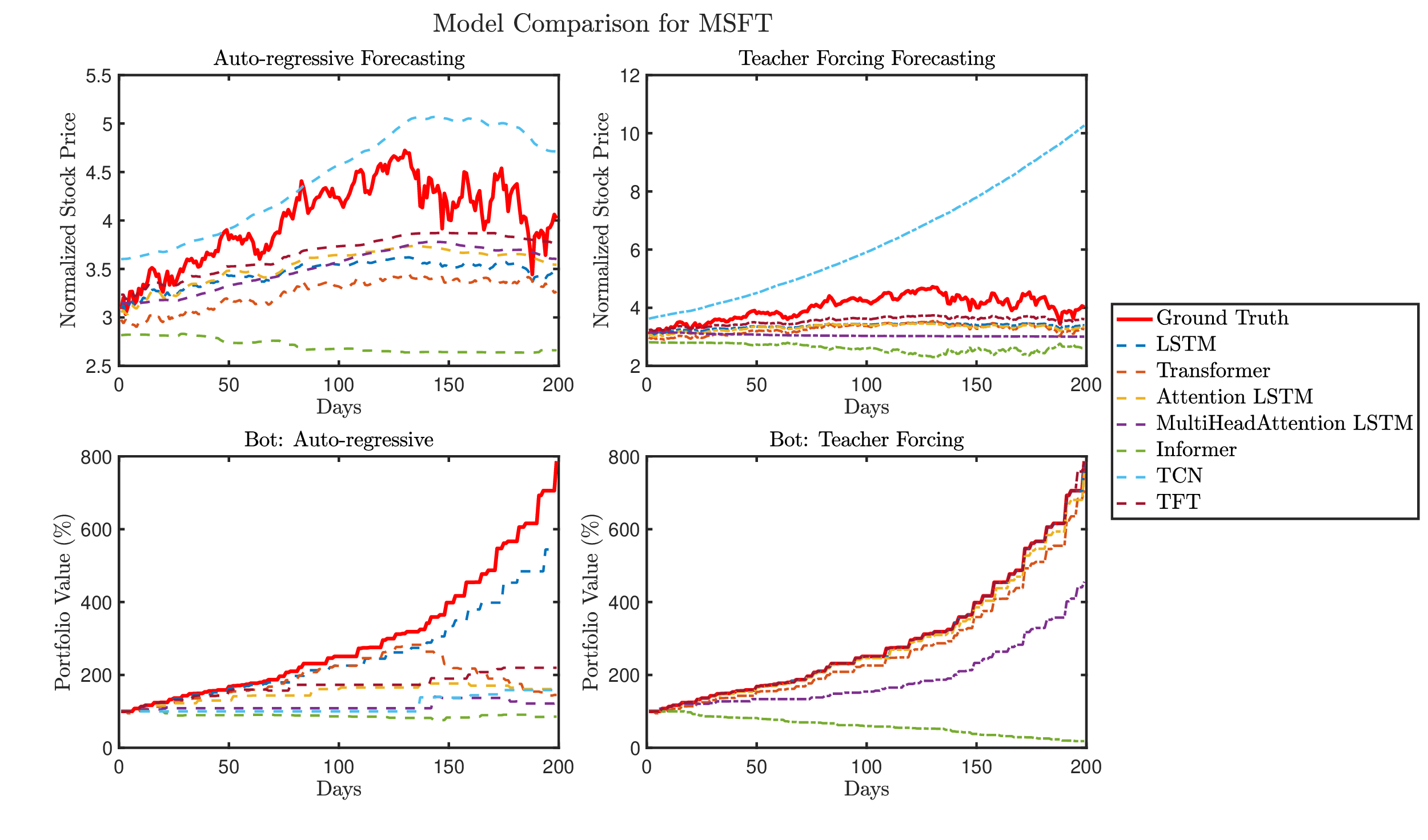}
\caption{One-day-ahead forecasting results for MSFT. Top-left: autoregressive price prediction. Top-right: teacher-forced price prediction. Bottom-left: portfolio value driven by autoregressive forecasts. Bottom-right: portfolio value driven by teacher-forced forecasts.}
\label{fig:msft_oneday}
\end{figure}

Figures~\ref{fig:aapl_oneday} and~\ref{fig:msft_oneday} summarize the results for stock ticker, AAPL and MSFT, respectively. In each figure, the top-left panel shows autoregressive price forecasts, while the top-right panel shows teacher-forced forecasts. The bottom-left panel reports the evolution of portfolio value when trading decisions are driven by autoregressive predictions, and the bottom-right panel shows the corresponding portfolio trajectory under teacher forcing.

Across both stocks, all models capture short-term trends, or at least directional trends under teacher forcing; however, notable differences emerge under autoregressive evaluation. The vanilla LSTM consistently exhibits the most stable long-horizon behavior, with slower error accumulation and reduced drift relative to attention-based and convolutional architectures. These stability advantages directly translate into smoother portfolio trajectories and fewer spurious trades. Specifically for the StockBot, a quantitatively accurate prediction is not a prerequisite since it depends only on the directional movement of the stock prices.

More expressive architectures such as transformers, TFT, and Informer models often achieve lower training loss, but exhibit increased sensitivity to compounding prediction errors when recursively deployed. In the absence of hyperparameter tuning, this sensitivity leads to degraded trading performance despite comparable short-term predictive accuracy.
\begin{table}[H]
\centering
\begin{tabular}{|l|cc|cc|}
\hline
 & \multicolumn{2}{c|}{\textbf{AAPL}} & \multicolumn{2}{c|}{\textbf{MSFT}} \\
\textbf{Model} 
& {Auto. RMSE} & {TF RMSE} 
& {Auto. RMSE} & {TF RMSE} \\
\hline
LSTM                     & 0.2556 & 0.3061 & 0.6129 & 0.7040 \\
Transformer              & 0.3713 & 0.3094 & 0.7952 & 0.7737 \\
Attention LSTM           & 0.5885 & 0.4732 & 0.5314 & 0.7422 \\
MultiHeadAttention LSTM  & 0.6137 & 0.5164 & 0.5502 & 1.0478 \\
Informer                 & 0.7670 & 0.7749 & 1.3772 & 1.4819 \\
TCN                      & 0.5805 & 0.7188 & 0.5295 & 2.8410 \\
TFT                      & 0.6587 & 0.6955 & 0.4254 & 0.5254 \\
\hline
\end{tabular}
\caption{Test RMSE comparison across forecasting modes for AAPL and MSFT. Autoregressive (Auto.) and teacher forcing (TF) results are reported for each model.}
\label{tab:rmse_comparison_combined}
\end{table}

Table~\ref{tab:rmse_comparison_combined} quantitatively summarizes test RMSE values for both forecasting modes, confirming that the LSTM achieves the lowest or near-lowest error under autoregressive and teacher-forcing deployment.

\subsection{Ten-Day-Ahead Forecasting}

We next evaluate a more challenging setting in which models predict a vector of ten future trading days simultaneously. This formulation reduces recursive feedback but places greater demands on the model’s ability to capture medium-range temporal structure.

\begin{figure}[H]
\centering
\includegraphics[width=0.95\textwidth]{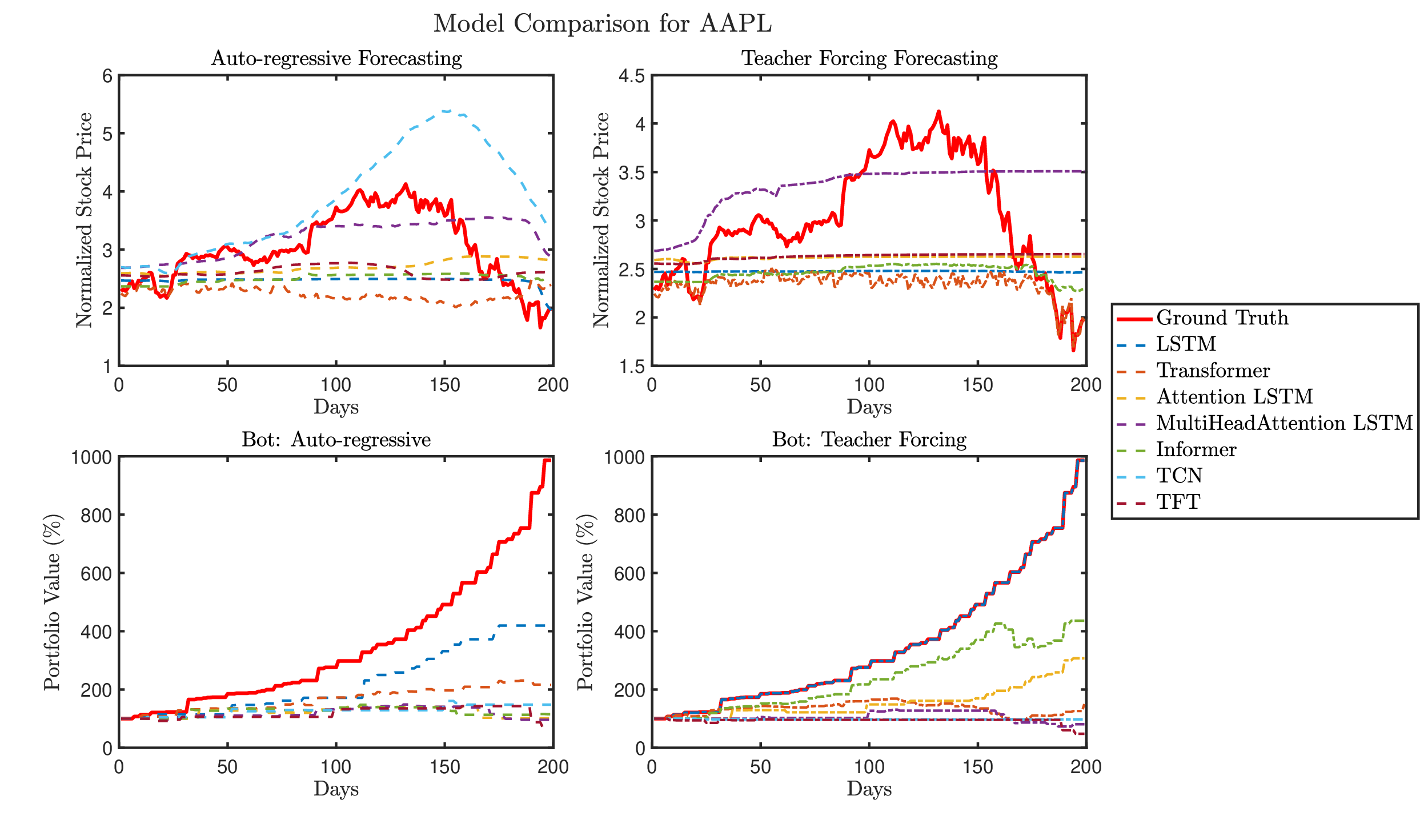}
\caption{Ten-day-ahead forecasting results for AAPL. Top-left: autoregressive price prediction. Top-right: teacher-forced price prediction. Bottom-left: portfolio value driven by autoregressive forecasts. Bottom-right: portfolio value driven by teacher-forced forecasts.}
\label{fig:aapl_tenday}
\end{figure}

\begin{figure}[H]
\centering
\includegraphics[width=0.95\textwidth]{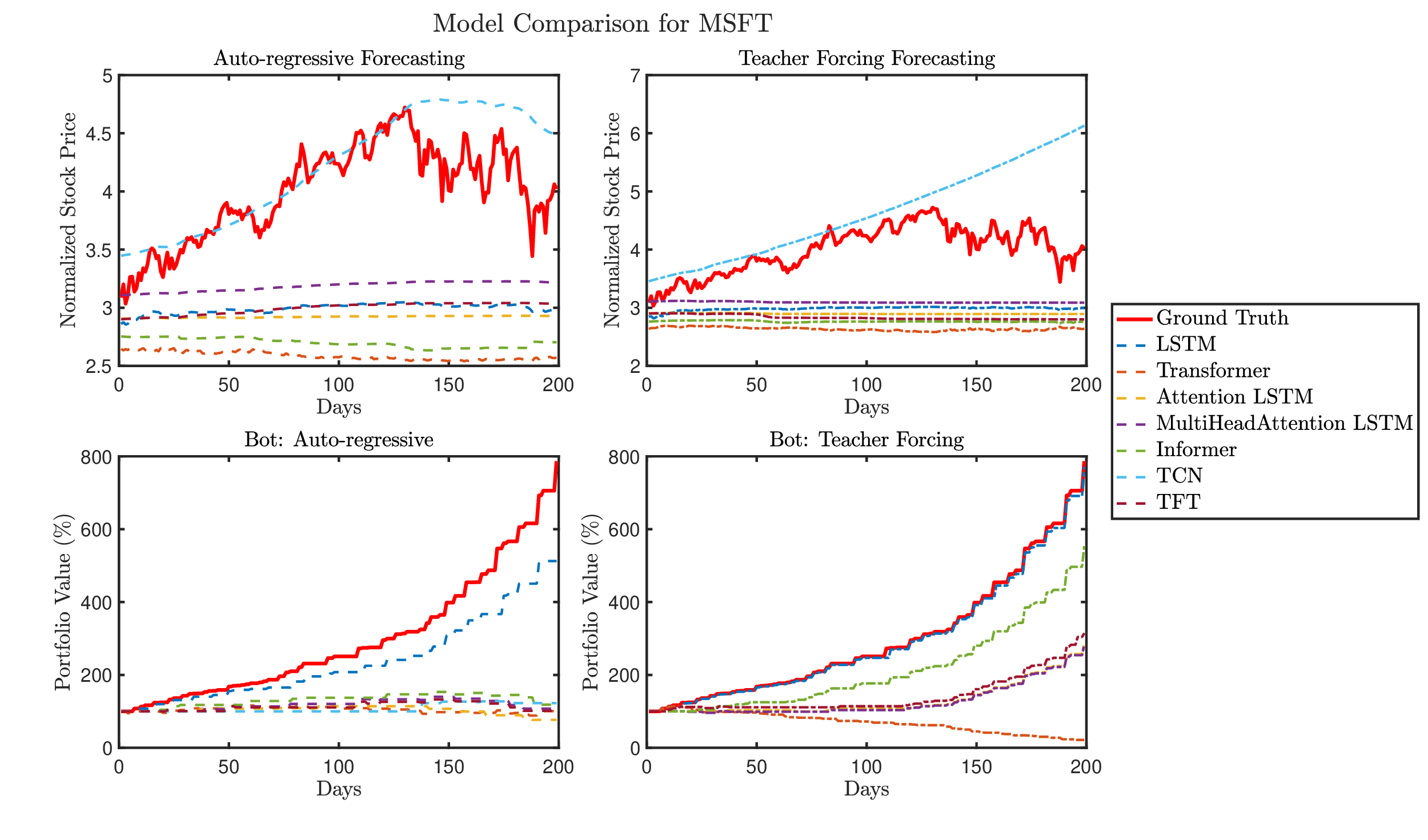}
\caption{Ten-day-ahead forecasting results for MSFT. Top-left: autoregressive price prediction. Top-right: teacher-forced price prediction. Bottom-left: portfolio value driven by autoregressive forecasts. Bottom-right: portfolio value driven by teacher-forced forecasts.}
\label{fig:msft_tenday}
\end{figure}

Figures~\ref{fig:aapl_tenday} and~\ref{fig:msft_tenday} present the corresponding results for stock tickers, AAPL and MSFT, respectively. As before, the top-left and top-right panels show autoregressive and teacher-forced price predictions, respectively, while the bottom panels report portfolio growth driven by the two forecasting strategies.

\begin{table}[H]
\centering
\begin{tabular}{|l|cc|cc|}
\hline
 & \multicolumn{2}{c|}{\textbf{AAPL}} & \multicolumn{2}{c|}{\textbf{MSFT}} \\
\textbf{Model} 
& {Auto. RMSE} & {TF RMSE} 
& {Auto. RMSE} & {TF RMSE} \\
\hline
LSTM                     & 0.8033 & 0.8190 & 1.0722 & 1.0845 \\
Transformer              & 1.0405 & 0.8749 & 1.4791 & 1.4319 \\
Attention LSTM           & 0.6997 & 0.7276 & 1.1506 & 1.1778 \\
MultiHeadAttention LSTM  & 0.5624 & 0.6294 & 0.8965 & 0.9958 \\
Informer                 & 0.7676 & 0.7812 & 1.3725 & 1.3022 \\
TCN                      & 1.1186 & 150.3449 & 0.3543 & 0.8802 \\
TFT                      & 0.7160 & 0.7167 & 1.0699 & 1.2448 \\
\hline
\end{tabular}
\caption{Test RMSE comparison across forecasting modes for AAPL and MSFT. Autoregressive (Auto.) and teacher forcing (TF) results are reported for each model.}
\label{tab:rmse_comparison_combined_multiday}
\end{table}

Compared to one-day-ahead prediction, all models exhibit increased uncertainty at longer horizons. Transformer-based and attention-augmented architectures produce smoother multi-day trajectories under teacher forcing, but their autoregressive deployment remains susceptible to drift when trained with default hyperparameters. In contrast, the LSTM maintains competitive accuracy and demonstrates robust trading behavior, yielding portfolio growth curves that are comparable or superior to more complex models, as qunatified in Table~\ref{tab:rmse_comparison_combined_multiday} which outlines the lower RMSE of the LSTM model under autoregressive and teacher-forecasting mode deployment.

These results highlight an important practical insight: while attention-based architectures offer greater modeling flexibility, strong recurrent inductive bias and data efficiency make vanilla LSTMs particularly well-suited for financial time-series forecasting in data-limited settings. When evaluation emphasizes recursive deployment and downstream decision-making, architectural simplicity and stability can outweigh raw expressiveness.
\section{Conclusions}
\label{sec:conclusions}

In this work, we conducted a systematic evaluation of recurrent, attention-based, convolutional, and hybrid temporal architectures for stock price forecasting within a unified experimental and decision-making framework. Contrary to common expectations derived from large-scale natural language and sequence modeling benchmarks, our results indicate that a carefully constructed vanilla LSTM consistently outperforms more sophisticated attention-driven models when trained under a common set of default hyperparameters and in data-limited financial settings.

Across both autoregressive and teacher-forcing evaluation modes, single-step (one-day-ahead) forecasting emerges as the most reliable regime for all model classes, yielding lower prediction error and more stable downstream trading behavior. While multi-day forecasting remains feasible, error accumulation and sensitivity to hyperparameter choices become increasingly pronounced for transformer-based and convolutional architectures, particularly in the absence of task-specific tuning. These findings highlight the importance of architectural inductive bias and optimization stability in financial time-series modeling, where data availability and signal-to-noise ratios differ fundamentally from large-scale sequence learning domains.

When embedded within the StockBot trading framework, LSTM-driven predictions translate into more consistent portfolio growth and reduced volatility relative to attention-based alternatives. Although transformer-style models offer greater representational flexibility and interpretability, their advantages do not manifest reliably under constrained data and uniform hyperparameter configurations. This suggests that model complexity alone is insufficient to guarantee superior performance in practical financial forecasting applications.

Overall, our results reaffirm the continued relevance of recurrent sequence models for market prediction and decision-making, particularly when robustness, data efficiency, and deployment simplicity are prioritized. For StockBot-specific applications, a more prudent approach to explore will be training forecasting classifiers instead of regressors, since buy/sell decisions are made purely on the basis of directional change.  Future work will focus on targeted hyperparameter optimization, hybrid recurrent–attention formulations, ensembling the models for uncertainty quantification, and regime-aware training strategies to better assess the conditions under which more expressive temporal models can provide tangible benefits over classical recurrent architectures. 



\section*{Data Access}
All datasets used in this study are publicly available through Yahoo Finance~\cite{Yahoofin} and Google Trends~\cite{GTAB}. Additional information on computation and code implementation can be obtained from the author upon request.

\section*{Acknowledgements}
The author acknowledges insightful discussions and feedback from colleagues and mentors that improved the clarity of this manuscript.

\bibliographystyle{unsrtnat}
\bibliography{reference}
\end{document}